\documentclass[twocolumn]{aastex631}

\usepackage{amsmath,amsfonts,amssymb}
\usepackage{graphicx}
\usepackage{xspace}
\usepackage{hyperref}
\usepackage{csquotes}

\renewcommand{\deg}{\textdegree\xspace}

\begin{document} 

\title{Using AI for Wavefront Estimation with the Rubin Observatory Active Optics System}

\correspondingauthor{John Franklin Crenshaw}
\email{jfc20@uw.edu}

\author[0000-0002-2495-3514]{John Franklin Crenshaw}
\affiliation{Department of Physics, University of Washington, Seattle, WA 98195, USA}
\affiliation{DIRAC Institute, University of Washington, Seattle, WA 98195, USA}

\author[0000-0001-5576-8189]{Andrew J. Connolly}
\affiliation{DIRAC Institute, University of Washington, Seattle, WA 98195, USA}
\affiliation{Department of Astronomy, University of Washington, Seattle, WA 98195, USA}
\affiliation{eScience Institute, University of Washington, Seattle, WA 98195, USA}

\author[0000-0002-2308-4230]{Joshua E. Meyers}
\affiliation{Kavli Institute for Particle Astrophysics and Cosmology, Stanford, CA 94305, USA}
\affiliation{SLAC National Accelerator Laboratory, Menlo Park, CA 94025, USA}

\author[0000-0002-6825-5283]{J. Bryce Kalmbach}
\affiliation{DIRAC Institute, University of Washington, Seattle, WA 98195, USA}
\affiliation{Department of Astronomy, University of Washington, Seattle, WA 98195, USA}

\author[0000-0001-6013-1131]{Guillem Megias Homar}
\affiliation{Kavli Institute for Particle Astrophysics and Cosmology, Stanford, CA 94305, USA}
\affiliation{SLAC National Accelerator Laboratory, Menlo Park, CA 94025, USA}
\affiliation{Department of Aeronautics and Astronautics, Stanford University, Stanford, CA 94305, USA}

\author[0000-0002-0138-1365]{Tiago Ribeiro}
\affiliation{Vera C. Rubin Observatory, Tucson, AZ 85719, USA}

\author[0000-0002-9589-1306]{Krzysztof Suberlak}
\affiliation{DIRAC Institute, University of Washington, Seattle, WA 98195, USA}
\affiliation{Department of Astronomy, University of Washington, Seattle, WA 98195, USA}

\author[0000-0002-9121-3436]{Sandrine Thomas}
\affiliation{Vera C. Rubin Observatory, Tucson, AZ 85719, USA}

\author[0009-0007-5732-4160]{Te-Wei Tsai}
\affiliation{Vera C. Rubin Observatory, Tucson, AZ 85719, USA}

\begin{abstract}
    The Vera C. Rubin Observatory will, over a period of 10 years, repeatedly survey the southern sky.
    To ensure that images generated by Rubin meet the quality requirements for precision science, the observatory will use an Active Optics System (AOS) to correct for alignment and mirror surface perturbations introduced by gravity and temperature gradients in the optical system.
    To accomplish this Rubin will use out-of-focus images from sensors located at the edge of the focal plane to learn and correct for perturbations to the wavefront.
    We have designed and integrated a deep learning model for wavefront estimation into the AOS pipeline.
    In this paper, we compare the performance of this deep learning approach to Rubin's baseline algorithm when applied to images from two different simulations of the Rubin optical system.
    We show the deep learning approach is faster and more accurate, achieving the atmospheric error floor both for high-quality images, and low-quality images with heavy blending and vignetting.
    Compared to the baseline algorithm, the deep learning model is 40x faster, the median error 2x better under ideal conditions, 5x better in the presence of vignetting by the Rubin camera, and 14x better in the presence of blending in crowded fields.
    In addition, the deep learning model surpasses the required optical quality in simulations of the AOS closed loop.
    This system promises to increase the survey area useful for precision science by up to 8\%.
    We discuss how this system might be deployed when commissioning and operating Rubin.
\end{abstract}


\section{Introduction}
\label{sec:intro}

The Vera C. Rubin Observatory's Legacy Survey of Space and Time (LSST) will spend ten years imaging the entire southern sky to unprecedented depth \citep{ivezic2019}.
LSST promises to improve cosmological constraints by an order of magnitude \citep{descSRD}, dramatically expand our understanding of galaxy evolution \citep{galaxiesSRM}, unveil the transient and variable sky \citep{tvsSRM}, and provide an inventory of the Solar System of unprecedented completeness \citep{ssscSRM}.

To enable the wide variety of science undertaken by LSST, the Rubin Observatory must deliver high and consistent optical quality across the entire 3.5 degree field of view (FoV).
The median seeing at Rubin's site on Cerro Panch\'on is 0.65" \citep{ivezic2019}, and Rubin's optical system is required to degrade this value by no more than 0.4" (see Appendix~\ref{sec:error-budget}).
To meet this requirement, the Rubin Observatory will use an active optics system (AOS) to provide real-time corrections to the optical alignment and mirror figure, which are perturbed by gravity and temperature gradients \citep{thomas2017, neill2014}.
The AOS uses 228 actuators to control the surface figure of the mirrors (156 actuators on the joint primary/tertiary mirror, called M1M3, and 72 actuators on the secondary mirror, called M2), and 2 hexapods to adjust the positions and rotations of M2 and the camera.
A position- and temperature-dependent lookup table (LUT) is used in an open-loop to apply optimal mean corrections before each exposure.
These mean corrections are, however, unable to account for hysteresis, dome seeing, small-scale temperature gradients, and temperature uncertainties.

The Rubin AOS will use a closed-loop to detect and correct these residual errors.
At each corner of the focal plane, there are curvature wavefront sensors (CWFSs) consisting of an offset pair of CCDs: one 1.5~mm inside of focus (intra-focal) and one 1.5~mm outside of focus (extra-focal), corresponding to 24~$\mu m$ \textbf{RMS} of wavefront defocus.
These out-of-focus images are used to infer perturbations to the optical wavefront, from which actuator and hexapod corrections are determined.
Due to the fast cadence of LSST, wavefront inference and correction of the optical system must be completed in less than 12 seconds, and this whole process must be repeated every 36 seconds for the duration of an observing run.
To deliver the optical quality required for LSST, the AOS must limit the PSF (Point Spread Function) FWHM (Full Width Half Maximum) contribution of optical aberrations to no more than 0.09", including only 0.079" due to misestimation of the wavefront -- less than 10\% of the total error budget (see Appendix~\ref{sec:error-budget}).

There are several features of Rubin's design that make wavefront estimation difficult:
(i) the primary mirror has a large central obscuration ($R_\text{inner}/R_\text{outer} = 0.61$), which results in a large loss of information about the wavefront;
(ii) the optical beam is very fast ($f/1.23$), which results in a non-linear projection of the reference sphere onto the pupil;
(iii) the CWFSs are at the edge of Rubin's wide field of view, generating pupil distortion and vignetting of the out-of-focus images;
(iv) the images from the intra- and extra-focal sensors are of different sources, which have different intrinsic properties and field-dependent wavefront variations that must be accounted for.

The baseline algorithm for wavefront inference is an iterative solver of the transport of intensity equation (TIE), described in \citet{xin2015}.
This algorithm has been implemented and tested in the Rubin AOS pipeline \citep{thomas2022a}, but there are several limitations that could be improved upon, including degradation of wavefront estimation in the presence of vignetting from the camera body and blending in crowded fields, and slow evaluation times.
The vignetting and blending limitations constrain source selection for wavefront estimation, and in the densest 8\% of fields it is expected that there will be no sources available for reliable wavefront estimation with the TIE.
The limitation on time reduces the number of sources that can be used to constrain the wavefront.

In this paper, we build, integrate, and test a deep learning (DL) wavefront estimator that uses convolutional neural nets (CNNs) to learn a mapping from individual out-of-focus images and metadata to the optical wavefront.
Previous papers have built prototype CNNs for wavefront estimation with Rubin \citep{thomas2020a, thomas2021}, but this is the first time such a system has been integrated into the Rubin AOS pipeline, and validated against the baseline algorithm.
In addition, \citet{yin2021} built a prototype network that directly estimates control parameters, without the need for first estimating the wavefront.
In this paper, we focus on estimating the wavefront, as this allows direct comparison to the baseline Rubin algorithm.
We show that our system outperforms the baseline algorithm on simulations of Rubin images, including dramatic improvements in the case of vignetting and blending, and faster evaluation times.

In Section~\ref{sec:sims} we describe the simulations we use to train and test our system.
In Section~\ref{sec:wfest} we describe the baseline wavefront estimation algorithm, and the details of our deep learning approach.
In Section~\ref{sec:results} we compare the performance of our deep learning system to the baseline algorithm.
We discuss these results and conclude in Section~\ref{sec:discussion}.

\section{Simulated data}
\label{sec:sims}

In this section, we describe the simulations used to train and test the deep learning model.
In particular, we train the network on 256\,000 donuts simulated with the Batoid \citep{batoid} package, and simulate validation and test sets each containing 32\,000 donuts.
The validation set is used to avoid overfitting during training, and the test set is used for computing metrics for comparison to the baseline wavefront estimation algorithm.
We also simulate the AOS closed loop using an independent raytracing package PhoSim \citep{phosim}, which allows us to test whether the network has overfit on details specific to the Batoid simulation.

\subsection{Batoid simulations}
\label{sec:batoid}

\begin{figure*}
    \centering
    \includegraphics{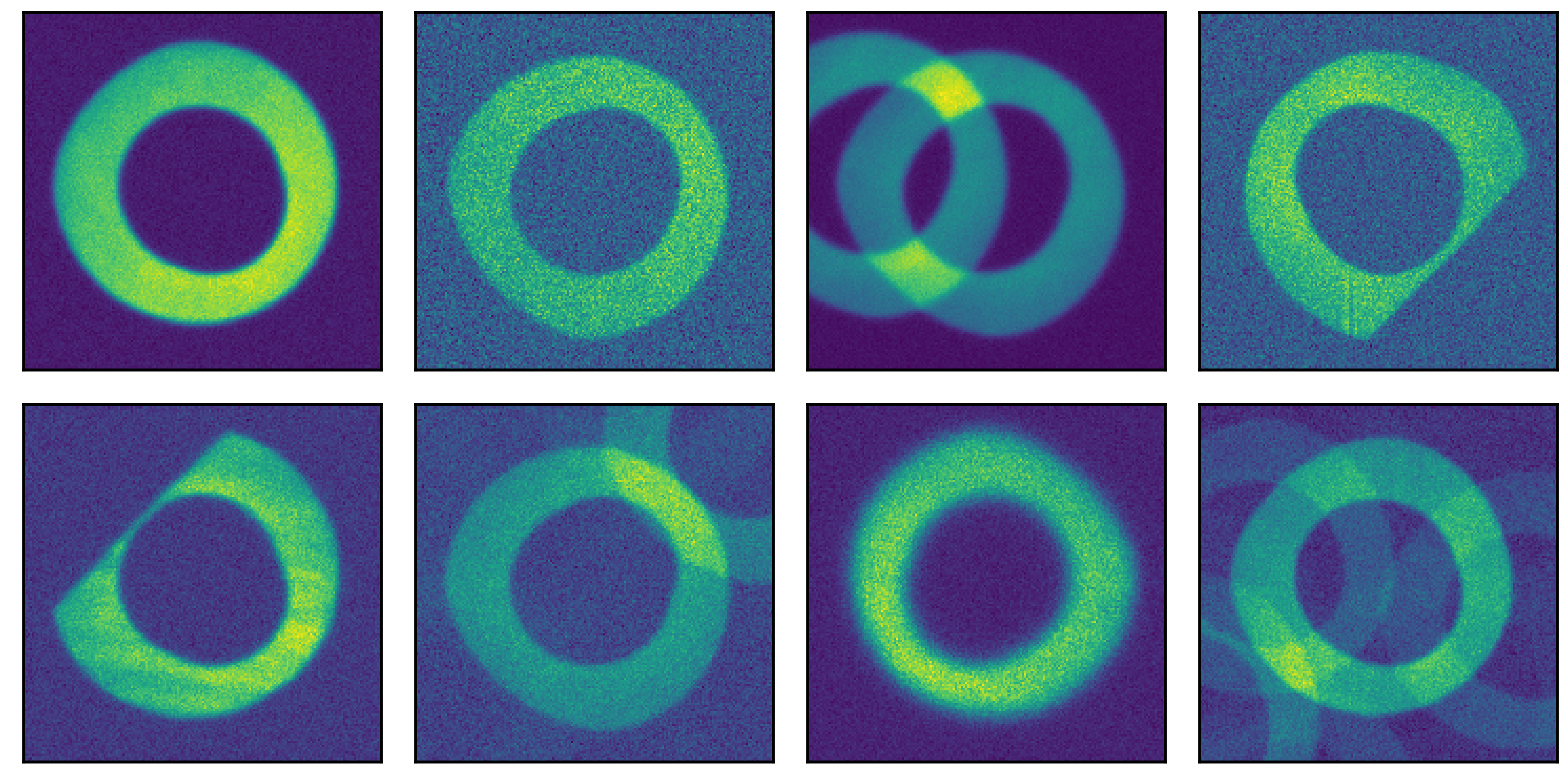}
    \caption{
        Example of donuts simulated with Batoid and Galsim.
        This set represents the diversity of the simulation:
        the simulated donuts have varying SNR, some are sharply vignetted by the Rubin camera (top right and bottom left), some are blended with a single other donut, and a small fraction are blended with many donuts.
    }
    \label{fig:donuts}
\end{figure*}

The network training data and the test data for the metrics in Section~\ref{sec:metrics} were generated using the Batoid \citep{batoid} raytracer to simulate the Rubin optical system.
The atmosphere, photometric bandpasses, and image sensors were simulated using GalSim \citep{galsim}.
The components of these simulations are detailed below:
\begin{enumerate}
    \item We select a set of observing conditions from the baseline LSST OpSim simulation \citep{opsim}. This includes the photometric band, the airmass, the 500~nm seeing at zenith, and the sky brightness. For details on the simulated atmosphere, see Appendix~\ref{sec:atm}.
    
    \item From the baseline OpSim simulation, we select a random astronomical pointing, and query the Gaia \citep{gaia} DR2 \citep{gaiaDR2} reference catalog for stars at this pointing that fall on the CWFSs. For each star, we retrieve the position, and the apparent Gaia magnitudes ($G$, $G_{BP}$, $G_{RP}$). We use the following equation from Chapter 5 of the Gaia DR2 documentation \citep{gaiaDR2docCh5} to estimate the equivalent magnitude in the SDSS $r$ band:
    \begin{align}
        \begin{split}
            r = G &+ 0.12879 - 0.24662 \cdot x \\ 
            &+ 0.027464 \cdot x^2 + 0.049465 \cdot x^3,
        \end{split}   
    \end{align}
    where $x = G_{BP} - G_{RP}$.
    We then assume the LSST $r$ magnitude is the same as SDSS, sample a random temperature in the range $4 \cdot 10^3 K < T < 10^4 K$, and use a blackbody spectrum to calculate the observed magnitude in the given LSST band. We clip bright magnitudes at 14 to save computational time during photon raytracing.
    
    \item We generate a random set of perturbations to the Rubin optical state, and  extract the Zernike coefficients of the corresponding wavefront perturbation. This is detailed in Appendix~\ref{sec:perturbations}.

    \item Given the set of stars and the estimated sky background we simulate these sources using Batoid and Galsim, with an exposure time of 15 seconds.

    \item We select the 8 brightest sources per CWFS and cut out stamps of 160x160 pixels, yielding 32 donuts per simulation.
\end{enumerate}
The slowest step is the atmosphere generation, so for each atmosphere generated in step 1, we repeat steps two through five 100 times.
We generate 100 different atmospheres, resulting in a total of 320\,000 donut stamps, which we split 80\%/10\%/10\% into training/validation/test sets.
Each of these sets have distinct sets of atmospheres and optical perturbations.

Fig.~\ref{fig:donuts} shows eight example donut stamps.
The shape of the donuts is due to their position at the far edge of the field of view, resulting in pupil distortion and vignetting of rays by the telescope and camera bodies.
For example, the left two donuts in the top row have soft vignetting from the edges of the telescope mirrors, while the top right and bottom left donuts have sharp vignetting from the camera body.
Some of these donuts have bright blends, which we define as an overlapping donut within 2 magnitudes of the central donut.
Others have faint blends.
The impact of these donut features on wavefront estimation will be explored in Section~\ref{sec:metrics}.

\subsection{PhoSim}
\label{sec:phosim}

To provide a test of the robustness of the neural network to the properties of the training sample (prior to real data being obtained by Rubin), we evaluate our Batoid-trained network on simulations from PhoSim \citep{phosim}, an independent raytracing simulation.
PhoSim simulates the atmosphere, telescope, camera, and detector, and was developed for the Rubin Observatory to model the expected images from the telescope.
PhoSim uses different simulation strategies from Batoid and GalSim, and produces donut images that are not identical to the Batoid donuts, both in their shapes and intensity patterns.
In addition, the PhoSim simulations we use have some features, such as mirror print through, that are not present in our Batoid simulations.
These features make the PhoSim simulations a good candidate for testing how well our model adapts to other data sets.
In Section~\ref{sec:closedLoop}, we use our deep learning model to estimate wavefronts as a part of the AOS closed loop simulated with PhoSim.

\section{Wavefront Estimation}
\label{sec:wfest}

The Rubin AOS uses out-of-focus images of stars, which appear as donuts, to estimate the wavefront of the telescope.
The wavefront, $\varphi$, is expressed as a linear combination of annular Zernike polynomials \citep{mahajan1981}, $Z_i$, with expansion coefficients
\begin{align}
    \alpha_i = \frac{1}{A} \int d^2\!\rho ~ Z_i(\rho) \varphi(\rho).
    \label{eq:coefficients}
\end{align}
$A$ is the area of the pupil, which implicitly defines the normalization of $Z_i$, and $\rho$ is a 2D position vector on the pupil.
The goal of wavefront estimation is to estimate the values of the coefficients $\alpha_i$.
Only coefficients 4-22 (numbered according to Noll's convention \citep{noll1976}) are currently used to define perturbations in Rubin's optical system \citep{xin2015}.

Note that even with perfect alignment and mirror figure, every telescope has non-zero off-axis optical aberrations, i.e. $\alpha_i^0 \neq 0$.
Since we do not aim to ``correct'' these intrinsic aberrations from the design of the telescope, we seek to instead estimate $\alpha_i - \alpha_i^0$.
We will continue using the symbol $\alpha_i$ and the term ``wavefront'', leaving the subtraction of the intrinsic aberrations implicit.
Furthermore, as the wavefront varies across the focal plane, we aim to estimate the wavefront at the center of the boundary between the intra- and extra-focal chip of each CWFS pair.
After estimating the wavefront, we scale each coefficient $\alpha_i$ by a factor $s_i$, such that $|s_i \alpha_i|$ is approximately the degradation of the PSF FWHM due to the excitation of Zernike mode $Z_i$ (see Appendix~\ref{sec:psf-conversion} for details).
For the rest of this paper, we use this scaling to express Zernike amplitudes and wavefront estimation errors in terms of the equivalent PSF FWHM degradation.

The baseline wavefront estimation algorithm of the Rubin AOS is explained in Section~\ref{sec:tie}.
In Section~\ref{sec:dl}, we discuss our deep learning model, including the design and training of the neural network.

\subsection{Baseline wavefront estimation with the transport of intensity equation}
\label{sec:tie}

The baseline wavefront estimation algorithm of the Rubin AOS uses the transport of intensity equation (TIE),
\begin{align}
    \nabla_\rho \cdot [I(\rho, 0) \nabla_\rho \varphi(\rho)] +
    k \, \frac{\partial I(\rho, z)}{\partial z}|_{z=0} = 0.
\end{align}
$I(\rho, 0)$ is the intensity pattern on the pupil, which is located at $z=0$.
The TIE is essentially a conservation of energy equation in the paraxial limit, and relates the $z$-derivative of the intensity to the curvature of the wavefront.
The AOS uses a pair of intra- and extra-focal images to approximate the intensity at the pupil and the $z$-derivative across the pupil.
The TIE is then solved using an expansion of the wavefront in annular Zernike polynomials\footnote{
    The solution can be found either by projecting onto the annular Zernike basis, or via Fast Fourier Transform (FFT).
    For this work, we use the projection method.
}.

The Rubin AOS does not take images of the same stars on both sides of focus, but rather relies on simultaneous images of different stars on distinct intra- and extra-focal sensors.
This formally violates the assumptions of the TIE, and requires additional steps to mitigate problems including different pupil masks and intensities, and variations of the wavefront across the focal plane.
The fast $f$-number, and the location of the CWFSs at the very edge of the field of view add additional complications.
\citet{xin2015} explains the algorithm in detail.
The existing implementation of this TIE solver in the Rubin AOS pipeline will serve as the baseline against which our deep learning model is compared.

\subsection{Deep learning wavefront estimation}
\label{sec:dl}

\begin{figure*}
    \centering
    \includegraphics[width=\textwidth]{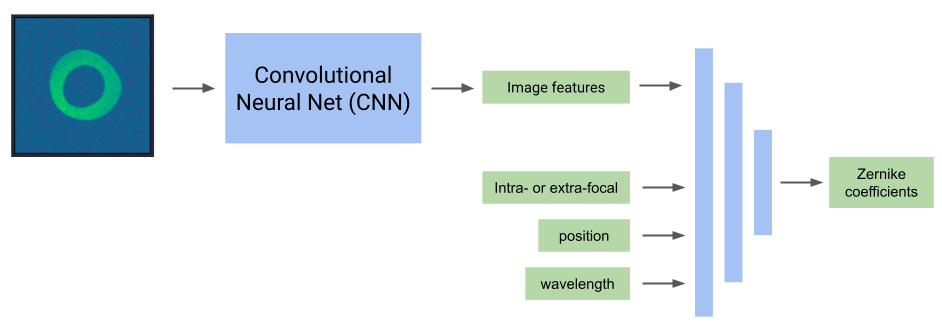}
    \caption{
        Schematic of the network architecture.
        A single donut is passed through a convolutional neural network (CNN), generating a set of image features.
        These features are concatenated with a flag that indicates if the donut is intra-/extra-focal, the field angle of the source, and the effective wavelength of the observation.
        A dense neural network then estimates the Zernike coefficients associated with the optical wavefront.
    }
    \label{fig:network}
\end{figure*}

The deep learning (DL) wavefront estimator is summarized in Fig.~\ref{fig:network}.
A convolutional neural network (CNN) extracts information from a single donut image, and produces a set of image features.
These features, along with some metadata, are fed into a set of fully connected layers that estimate the Zernike coefficients of the wavefront.
The metadata consists of:
(i) a flag indicating whether the image is intra- or extra-focal;
(ii) the field angle of the star;
(iii) the effective wavelength of the photometric band in which the donut was imaged.

We test two different architectures, both of which have the structure presented in Fig.~\ref{fig:network}, but differ in the details of the network layers contained inside the CNN and fully connected stages.
The first is the architecture designed for wavefront estimation by \citet{thomas2021}.
The second uses ResNet-18 \citep{resnet} (with the final prediction layer removed) as the CNN, plus three fully connected layers to predict Zernike coefficients from the image features and metadata.
ResNet-18 is a network that was trained on image classification for the ImageNet Large Scale Visual Recognition Challenge (ILSVRC).
We initialize ResNet-18 with its pretrained parameters, and test versions where those parameters are frozen during training, and where those parameters are free to change with the rest of the network parameters.
For the fully connected layers, we use three layers of size (171, 57, 19), the first two of which are followed by batch normalization and ReLU activation.
The output dimension 19 corresponds to the 19 Zernike coefficients we are estimating.
The former two dimensions were chosen to be equally spaced logarithmically between 516 (512 image features output by ResNet-18 $+$ 4 items of metadata) and the output dimension (19).

For both network architectures, we apply the same preprocessing to the data.
We rescale each image so that all pixel values are between 0 and 1.
We then calculate the mean and variance of all pixels in the training images, and whiten the pixels 
\begin{align}
    \text{pixel} \to \frac{\text{pixel} - \text{mean}}{\sqrt{\text{variance}}}.
\end{align}
We also calculated the training-set mean and variance of each item of metadata, and whitened them as well.
Note that these same transformations are applied to the test set before network evaluation, using the means and variances calculated from the training set.

We train the networks to minimize the residual sum of squares (RSS) of the estimated Zernike coefficients and the true coefficients, including the weighting for the PSF FWHM degradation.
In other words, the network is trained to minimize the PSF width.
We begin training using Adam \citep{adam} with a learning rate of $3 \times 10^{-4}$, and decrease the learning rate by a factor of 10 whenever the validation loss does not decrease for 10 epochs.
We stop training whenever the validation loss has not improved for 20 epochs.

When training the network whose CNN is the pre-trained ResNet-18, we have a choice to make about how to handle the pre-trained parameters.
We can keep the pre-trained parameters fixed during training, and only allow the fully connected layers to adjust their parameters, or we can unfreeze ResNet-18, and allow its parameters to be tweaked during training as well.
We found that unfreezing the ResNet-18 parameters resulted in a much better validation loss.

Both the network from \citet{thomas2021} and ResNet-18 with unfrozen parameters achieved good validation losses, plateauing near the atmospheric error floor (see Section~\ref{sec:metrics-ideal} and Appendix~\ref{sec:atm}).
However, the network from \citet{thomas2021} did not achieve as high of an optical quality as the unfrozen ResNet-18 when tested in the simulated AOS closed loop (see Section~\ref{sec:closedLoop}).
This suggests it is not as adept at domain adaptation as ResNet-18.
For the rest of this paper, we consider only the model that uses ResNet-18 with unfrozen parameters.
We note that \citet{yin2021} tested a wider variety of neural network architectures for a similar AOS task, and also found that ResNet-18 achieved the best results.

\section{Results}
\label{sec:results}

\begin{figure*}[h!]
    \centering
    \includegraphics{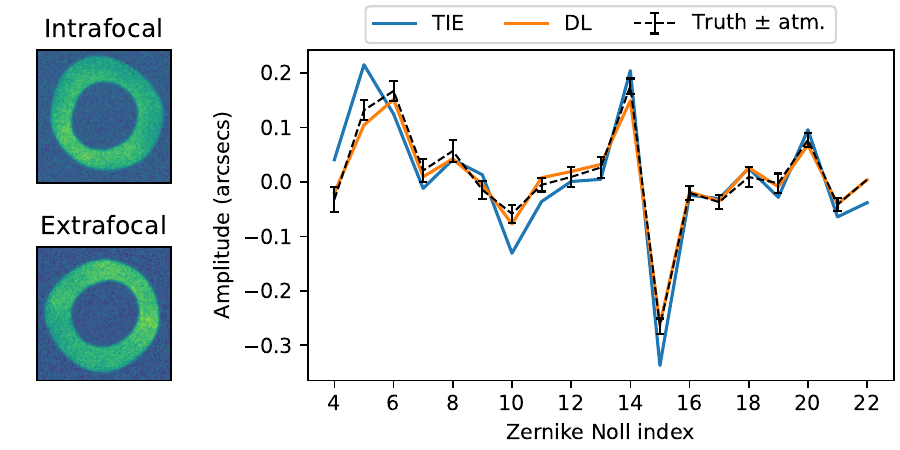}
    \caption{
        Example of an intra- and extra-focal donut fed into the AOS pipeline, with the DL and TIE wavefront estimates compared to the truth from the simulation.
        The plotted amplitudes are the quantities $s_i \alpha_i$ and $s_i \hat{\alpha}_i$ -- i.e., the contribution of each Zernike mode to the PSF FWHM.
        The error bars on the truth are the expected median $1\sigma$ errors due to the atmosphere.
    }
    \label{fig:zk-example}
\end{figure*}

In this section we compare the performance of the deep learning (DL) model to the baseline wavefront estimator (TIE).
First, we evaluate performance on the Batoid test set.
Then, we test both algorithms on the AOS closed loop simulated with PhoSim.
For both of these sections, we use the DL algorithm to estimate the wavefront from pairs of intra- and extra-focal donuts, and average the wavefront estimation from each.
This is because the TIE solver must operate on intra-/extra-focal pairs.
Averaging DL estimates for the same pairs of donuts allows an apples-to-apples comparison of the two algorithms.
In the final section, we evaluate the DL model on individual donuts, and discuss the trade-offs with estimation using pairs of donuts.

For all sections below, we define the error as the sum in quadrature of the PSF-weighted wavefront residuals,
\begin{align}
    \mathrm{Error} = \sqrt{\sum_{i=4}^{22} s_i^2 (\hat{\alpha_i} - \alpha_i)^2},
\end{align}
where $\hat{\alpha_i}$ is the estimated coefficient, and $\alpha_i$ is the truth from the simulation.
Due to the weights $s_i$, this error is approximately the degradation of the PSF FWHM in arcseconds due to the misestimation of the wavefront (see Appendix~\ref{sec:psf-conversion}).

We note that it takes about 70~ms for the AOS pipeline to return a wavefront estimate when using our DL model, while it takes more than three seconds when using the TIE solver.
Thus the DL algorithm is about 40 times faster than the baseline algorithm when estimating the wavefront from a single pair of donuts.
Even more substantial speed-ups are possible if you do not constrain the neural network to operate on pairs one-at-a-time like the TIE solver does, as neural networks are very efficient at massive, parallel evaluation.
In other words, the DL model enables vectorized wavefront estimation.\footnote{Note you can parallelize the TIE solver by sending each donut pair to a different processor, but then the number of donuts you can process is limited by the number of processors available for parallel computation.}
This speed comes at the up-front computational cost of simulations and network training, but given the goal is to design an algorithm that is able to operate quickly during real-time operations, this trade-off is worthwhile.

\subsection{Performance on Batoid test set}
\label{sec:metrics}

In this section, we evaluate the performance of the DL and baseline algorithms on pairs of donuts from the Batoid test set, first under ideal conditions, then with varying SNR, vignetting, and blending.

\subsubsection{Performance in ideal conditions \& comparison to expected atmospheric error}
\label{sec:metrics-ideal}

\begin{figure*}
    \centering
    \includegraphics{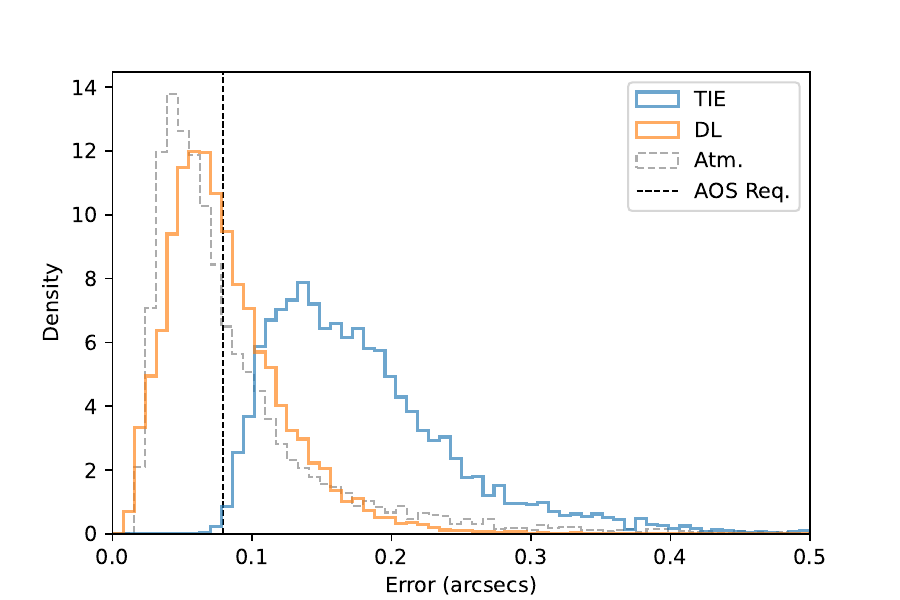}
    \caption{
        Distribution of wavefront estimation errors for the DL and TIE methods in ideal conditions (no camera vignetting and no bright blends).
        The distribution in gray is the expected distribution of wavefront estimation errors due to the phase variance of the atmosphere.
        The vertical black line is the AOS wavefront estimation error requirement of 0.079".
    }
    \label{fig:ideal}
\end{figure*}

First, we evaluate the performance on an idealized sample that contains no camera vignetting and no bright blends.
An example is shown in Fig.~\ref{fig:zk-example}.
An intra- and extra-focal donut are fed into the AOS pipeline, and both the DL and TIE algorithms are used to estimate the Zernike coefficients.
The estimates from each method are shown, along with the true wavefront from the simulations.
The error bars on the true wavefront are the median errors expected from the atmosphere (see Appendix~\ref{sec:atm}).
In ideal conditions, an optimal estimator should be consistent with the truth, within the uncertainty from the atmosphere.
In this example, the DL estimate is consistent with the truth within this uncertainty ($p\text{-value} \sim 0.89$), while the TIE estimate is not ($p\text{-value} \sim 0$).

To get a sense of how these methods perform on average, we plot the error distributions for each in Fig.~\ref{fig:ideal}.
The DL model systematically outperforms the TIE: the TIE median error is 2 times greater than the DL median, and the low-error tail of the TIE is still on the high-error side of the DL distribution.
The expected error distribution from the atmosphere alone is plotted in gray.
This distribution is very close to the DL error distribution, suggesting the DL errors are dominated by the irreducible atmospheric errors.
The TIE error distribution, on the the other hand, is shifted to significantly greater errors, indicating a model bias.
This is not surprising, as our TIE solver violates several assumptions of the basic TIE method (see Section~\ref{sec:tie}).

The AOS wavefront estimation error requirement \citep{LTS-186, LTS-124} of 0.079" is also marked by a vertical black line in Fig.~\ref{fig:ideal}.
The DL model is able to meet this requirement 57\% of the time using only a single pair of donuts.
In contrast, the TIE solver virtually never meets this requirement using only a single pair.
Note that you still need a wavefront solution from at least three of the CWFSs in order to interpolate the wavefront solution across the focal plane.
This carries the added benefit of averaging over the atmospheric errors that are correlated across any individual CWFS.

\subsubsection{Varying SNR}
\label{sec:metrics-snr}

\begin{figure*}
    \centering
    \includegraphics{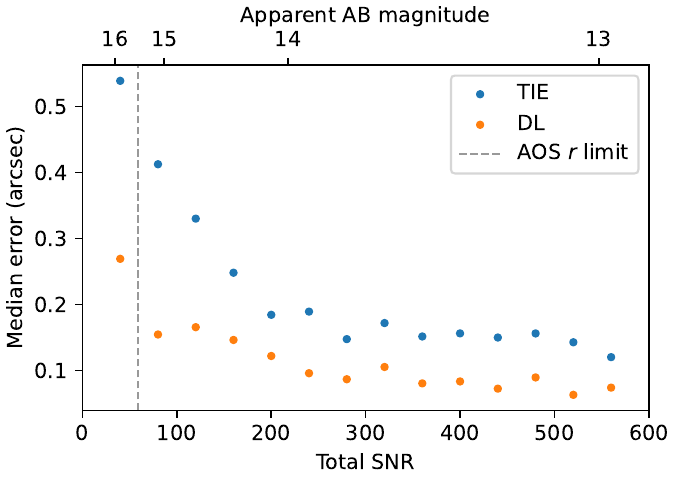}
    \caption{
        Median wavefront estimation error as a function of the donut SNR in the $r$ band.
        The x-axis below the plot shows the total SNR, while the x-axis above the plot shows corresponding $r$ band source magnitude, assuming the median $r$ band sky brightness of 21.20~AB~mag~arcsec$^{-2}$.
        The vertical gray line marks the current minimum SNR specified by the Rubin AOS pipeline.
    }
    \label{fig:snr}
\end{figure*}

To assess the impact of the image signal-to-noise ratio (SNR), we simulate pairs of stars (isolated, with no camera vignetting) imaged in the Rubin $r$ band with SNR ranging from 10 to 600.
We define the total SNR as the total source flux divided by the noise, where the noise includes both the Poisson uncertainty in the source flux and the sky background:
\begin{align}
    \mathrm{noise}^2 = \mathrm{flux} + N_\mathrm{pixels} \cdot \sigma_\mathrm{sky}^2,
\end{align}
where $N_\mathrm{pixels} = \pi R_d^2 (1 - \epsilon^2)$ is the number of pixels in a donut.
For Rubin, $R_d \approx 66$ pixels is approximately the radius of a donut, and $\epsilon=0.61$ is the fractional obscuration of the primary mirror.

Fig.~\ref{fig:snr} shows the median wavefront estimation error as a function of the SNR, along with the current minimum SNR specified by the Rubin AOS pipeline.
The top axis shows the corresponding source magnitude for the median $r$ band sky brightness of 21.20~AB~mag~arcsec$^{-2}$ \citep{ivezic2019}.
Both methods plateau at high SNR, corresponding to the medians of the distributions in Fig.~\ref{fig:ideal}.
Both methods begin to degrade once the total SNR drops below 200.
The DL model degrades more slowly, however, and its performance near the AOS SNR limit is similar to the performance of the TIE in the high-SNR regime.

\subsubsection{Camera vignetting}
\label{sec:metrics-vignetting}

\begin{figure*}
    \centering
    \includegraphics[width=\textwidth]{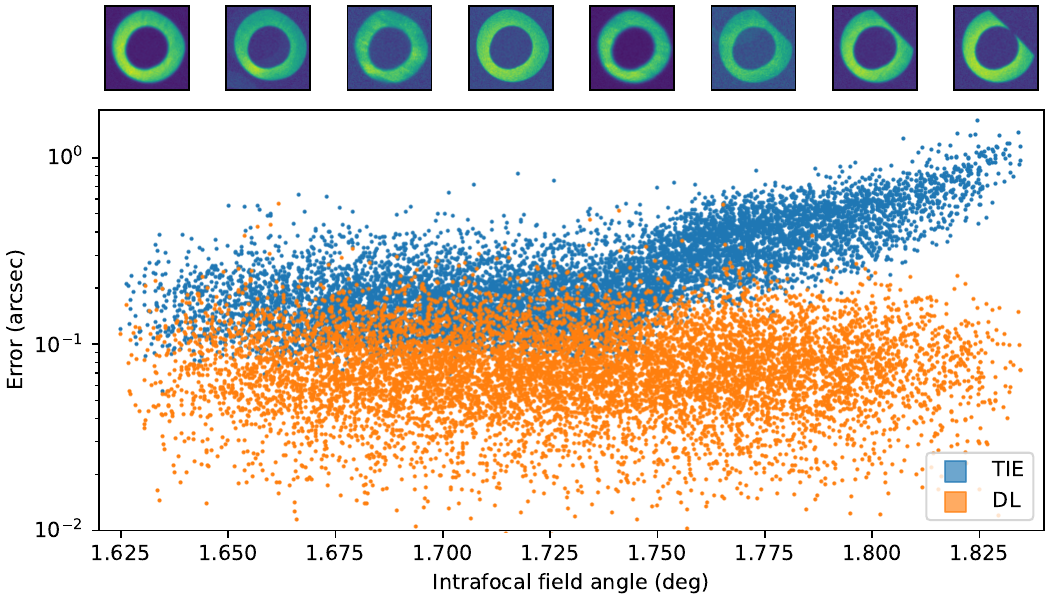}
    \caption{
        Wavefront estimation error as a function of intra-focal field angle.
        Above the scatter plot are donut stamps representative of the given field angle.
        Sharp vignetting from the camera body impacts field angles greater than about 1.74\deg.
        The data set plotted excludes donuts with bright blends.
    }
    \label{fig:vignetting-panel}
\end{figure*}

In this section, we investigate the impact of camera vignetting on wavefront estimation.
All of our donuts experience some amount of vignetting.
For donuts at a field angle less than about 1.74\deg, this is a soft vignetting by the mirror edges.
For donuts at a greater field angle, there is sharp vignetting from the body of the Rubin camera.
This sharp vignetting impacts 43\% of the intra-focal CWFSs, which are farther from the center of the focal plane than the extra-focal CWFSs.

\begin{figure*}[hbt!]
    \centering
    \includegraphics[width=\textwidth]{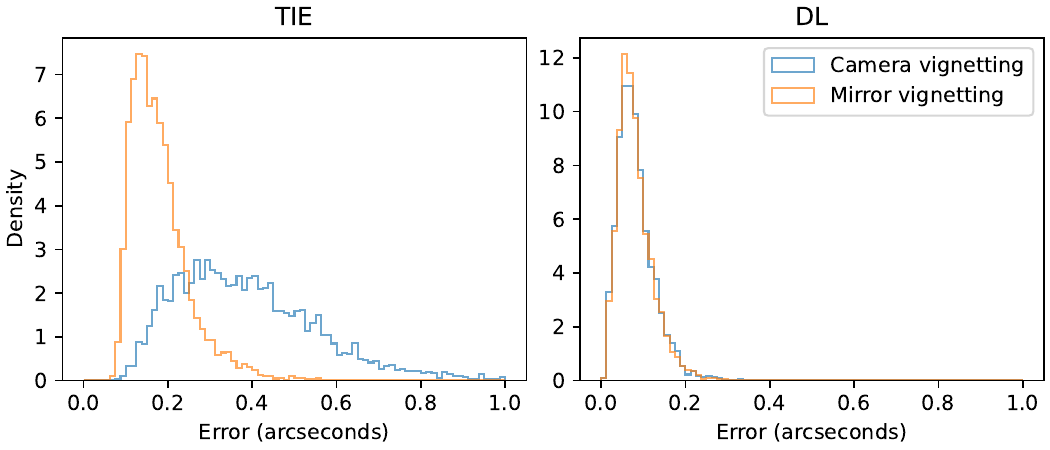}
    \caption{
        Left: distribution of errors for the TIE solver, in the presence of mirror and camera vignetting.
        Right: same for the DL model.
        The data set plotted excludes donuts with bright blends.
    }
    \label{fig:vignetting-hist}
\end{figure*}

In Fig.~\ref{fig:vignetting-panel}, you can see the wavefront estimation errors as a function the field angle of the intra-focal donut.
Above the scatter plot, you can see donut stamps characteristic of the given field angle.
Below a field angle of 1.74\deg, there is soft vignetting from the mirror edges, and beyond this point there is sharp vignetting from the camera body.
This transition is associated with a dramatic decline in the accuracy of wavefront estimation with the TIE solver.
On the other hand, the DL model is not appreciably impacted by camera vignetting.

This effect can also be seen in Fig~\ref{fig:vignetting-hist}, which shows the wavefront error distributions for these two regimes.
For the TIE, the camera-vignetted distribution is shifted towards higher errors, while for the DL model, the two distributions are indistinguishable.
For camera vignettes, the TIE median error is 5 times worse than the DL median.

\subsubsection{Blending}
\label{sec:metrics-blending}

\begin{figure*}
    \centering
    \includegraphics[width=0.9\textwidth]{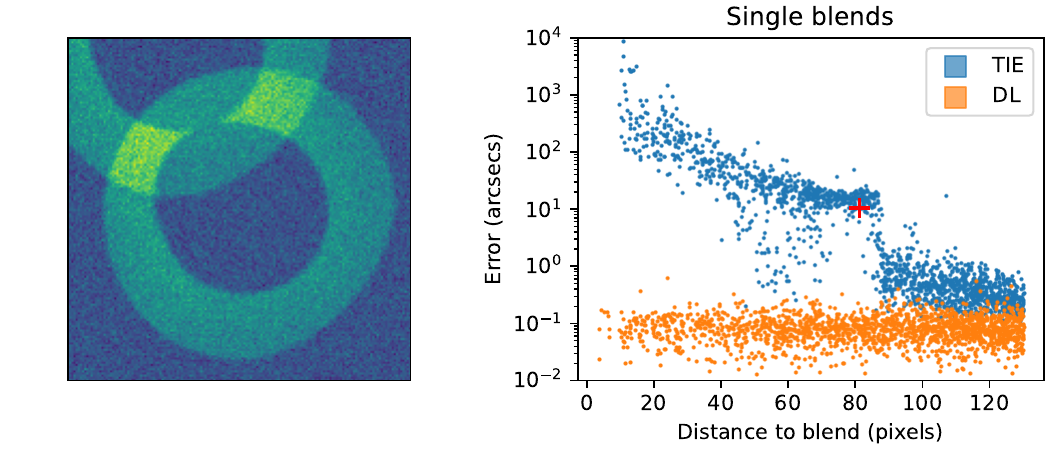}
    \caption{
        Left: a bright blend where the blending donut is at a distance of 82 pixels.
        Right: wavefront estimation error for donuts with a single bright blend, as a function of the centroid-to-centroid distance of the central and blending donut.
        The red cross denotes the distance and TIE error corresponding to the blend depicted on the left.
        The data set plotted excludes donuts vignetted by the camera.
    }
    \label{fig:blend-dist}
\end{figure*}

\begin{figure*}
    \centering
    \includegraphics[width=0.9\textwidth]{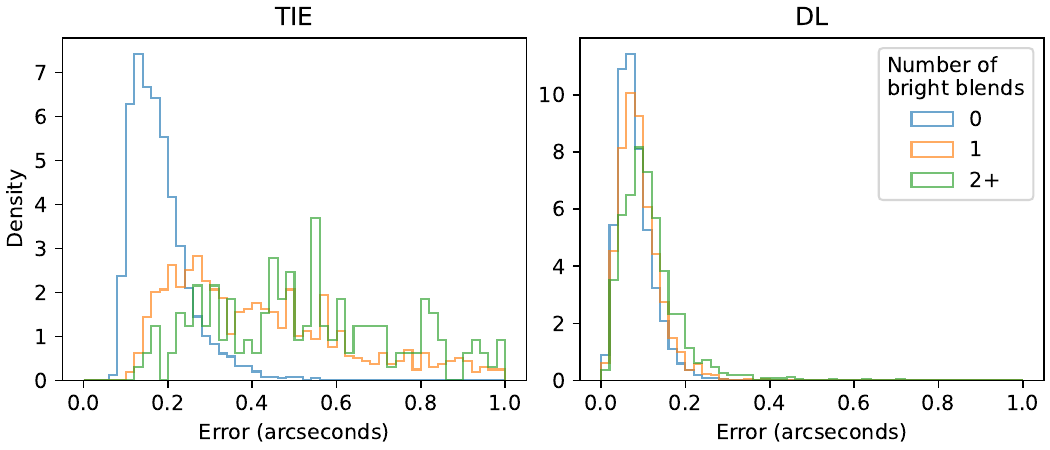}
    \caption{
        Left: distribution of errors for the TIE solver for different numbers of bright blends.
        Right: same for the DL model.
        The data set plotted excludes donuts vignetted by the camera, and blends for which the centroid-to-centroid distance is less than 88 pixels.
    }
    \label{fig:blend-bright}
\end{figure*}

\begin{figure*}
    \centering
    \includegraphics{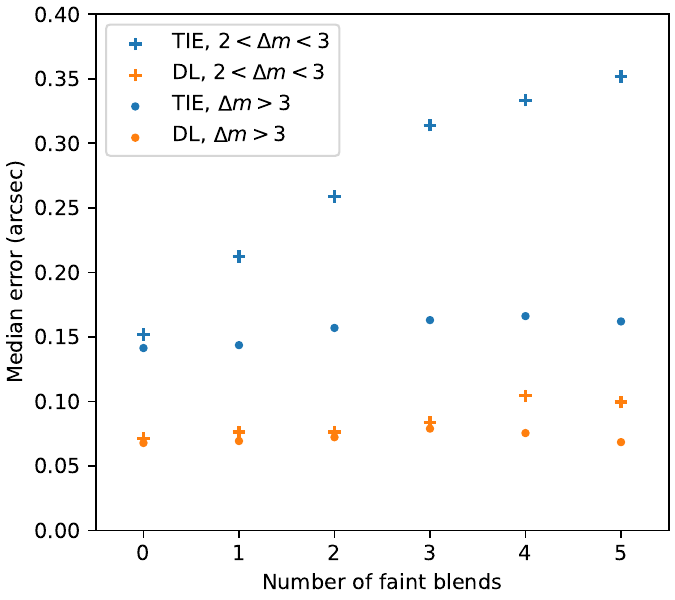}
    \caption{
        Median wavefront estimation error as a function of the number of faint blends.
        The faint blends are divided into two groups: those where the faint blends are at most 1 magnitude beyond the bright blend cut ($2 < \Delta m < 3$), and those where the faint blends are more than 1 magnitude beyond the bright blend cut ($\Delta m > 3$).
        The data set plotted excludes donuts vignetted by the camera, and donuts with bright blends.
    }
    \label{fig:blend-faint}
\end{figure*}

In this section, we evaluate the impact of blending on wavefront estimation.
First, we investigate bright blends, which we define as blends where the overlapping star is within 2 magnitudes of the central star.

Considering only donuts with a single bright blend, in Fig.~\ref{fig:blend-dist} we plot the wavefront estimation error vs the distance between the centroids of the central donut and the blending donut.
It is evident that the performance of the TIE solver degrades substantially once the blending donut is within approximately 88 pixels of the central donut (for reference, the donut radius is approximately 62 pixels).
The left panel of Fig.~\ref{fig:blend-dist} shows a blended donut for which the TIE estimate is substantially degraded, with the corresponding TIE error marked with a red cross on the right.
You can see the distance at which TIE performance substantially degrades roughly corresponds to the distance at which the central obscurations of the two donuts begin to overlap.
Whether the central obscurations for any specific pair of donuts actually overlap at this distance, however, is a function of field angle, orientation of the blend, and the optical perturbations.
Interestingly, whether or not the central obscurations actually overlap is not a deterministic predictor of whether the TIE catastrophically fails.
Regardless, the DL model appears robust to this transition, with similar performance in both regimes.

After removing donuts with bright blends at a distance less than 88 pixels, we compare the performance as a function of the number of bright blends.
In Fig.~\ref{fig:blend-bright}, you can see the TIE solver is significantly worse with a single bright blend.
With 2 or more bright blends, the TIE solver typically fails catastrophically, because the TIE solver masks blended pixels, and with more than a single blend, too much information is lost.
This results in the noisy distribution in Fig.~\ref{fig:blend-bright}, which has far fewer samples than the unblended and single blend cases. 
For the DL model, on the other hand, the single blend distribution overlaps the unblended distribution, and the 2+ blend distribution is only slightly degraded.
This demonstrates the robustness of the DL model to blending.
For single blends where the TIE is relatively successful, the TIE median error is still 14 times worse than the DL median.

Previously, we have only considered bright blends, which were defined as blends within 2 magnitudes of the central star.
However, fainter blends essentially modulate the background of the donut, and can also impact wavefront estimation.
We define $\Delta m$ to be the difference between the magnitude of the blending star and the central star, and divide the faint blends into two groups: those where the faint blends are at most 1 magnitude beyond the bright blend cut ($2 < \Delta m < 3$), and those where the faint blends are more than 1 magnitude beyond the bright blend cut ($\Delta m > 3$).

Fig.~\ref{fig:blend-faint} plots the median wavefront estimation error as a function of the number of both kinds of faint blends.
Both methods are robust to the presence of faint blends that are greater than 3 magnitudes fainter than the central star.
However, the TIE method degrades as you add blends between 2 and 3 magnitudes fainter than the central star (these are blends less than 1 magnitude past the bright blend cut).
The DL model also experiences some degradation, although to a lesser extent and with more faint blends.

These results indicate the DL model has the potential to dramatically improve AOS performance in very crowded fields where significant blending is unavoidable.
These results also provide information that can be used to improve the donut selection criteria for the TIE algorithm.

\subsection{Performance on PhoSim closed loop}
\label{sec:closedLoop}

\begin{figure*}
    \centering
    \includegraphics[width=\textwidth]{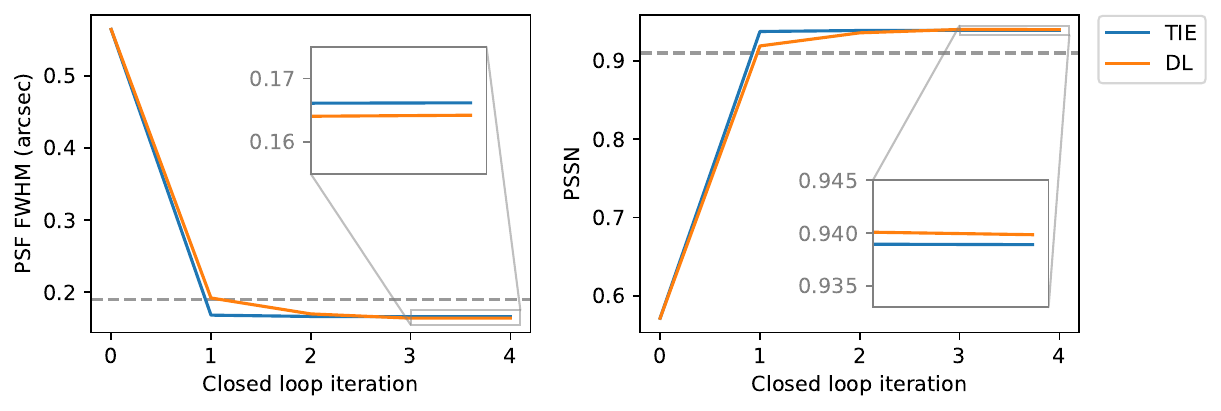}
    \caption{
        The PSF FWHM and PSSN as a function of closed loop iteration, when using the DL and TIE methods for wavefront estimation.
        The AOS requirements are marked by the horizontal dashed lines.
        The PSF FWHM must be below this limit, and the PSSN must be above this limit.
        \textbf{For both metrics, there is an inset that zooms into the last two iterations, where you can see the DL method slightly outperforms the TIE.}
    }
    \label{fig:closed-loop}
\end{figure*}

We wish to understand how the DL wavefront estimator performs when integrated into the full AOS pipeline, including the optical controller that takes the wavefront estimate as an input, and outputs commands to the actuators and hexapods.
We also wish to understand how the method performs on data sets not generated with Batoid, to ensure the DL model has not overfit on details specific to Batoid that do not translate to real data.
As we do not yet have real images from Rubin, we will perform this test on PhoSim simulations (see Section~\ref{sec:phosim} for details).

We use PhoSim to simulate the AOS closed loop.
That is, we start with a simulated telescope with random optical perturbations, and simulate unblended, unvignetted, high-SNR donut images on the intra- and extra-focal chips of each CWFS.
We cut out a 160x160 postage stamp of each donut, and use the DL algorithm to estimate the wavefront for each.
This information is fed into the optical feedback controller, which uses the wavefronts to derive optimal forces to apply to each actuator and hexapod, and these commands are applied to the simulated telescope.
The updated telescope model is then used to re-simulate images of the same stars, and this process is repeated in a loop.
If the estimated wavefront is close to the true wavefront, the optical quality will improve, and on the next iteration, the true wavefront will have smaller perturbations.
The estimate of the wavefront perturbations is expected to decrease with each iteration, until the simulated telescope is near perfect alignment and mirror figure.
It is important to test whether our DL estimator is capable of estimating wavefronts that drive the system towards and maintains the required optical quality.

The two metrics we track here are the PSF FWHM and the PSSN (Normalized Point Source Sensitivity).
The PSF FWHM represents the resolution of imaging, and PSSN is related to the relative SNR of observations and therefore represents the efficiency loss in telescope observing time \citep{seo2009}.
The AOS closed loop is required to converge to a FWHM of less than 0.19" and a PSSN greater than 0.91 (see Appendix~\ref{sec:error-budget}).

The PSF FWHM and PSSN as a function of closed loop iteration are plotted in Fig.~\ref{fig:closed-loop} for the DL and TIE algorithms.
In these simulations, both algorithms meet the requirements, which are plotted in gray.
Furthermore, after reaching their optimal values, both algorithms maintain the same optical quality on subsequent iterations, demonstrating the robustness of the DL model to atmospheric noise and the stability of the system when working together with the optical controller.
\textbf{
In the inset zoom-in's, you can see the DL model slightly outperforms the TIE on both metrics.
This improvement is modest, however, indicating the real strength of the DL method will not be in the case where you have an unblended, unvignetted, high-SNR donut on every wavefront sensor, but rather in the cases where one or multiple of these optimistic assumptions are broken.
Testing closed loop convergence in less optimal conditions is a future goal of the Rubin active optics team.
}

This test also demonstrates the robustness of the DL model to the differences between PhoSim and Batoid that are described in Section~\ref{sec:phosim}, and provides a check that training the DL model to optimize the PSF FWHM does not come at the expense of the PSSN (there is more discussion of this in Appendix~\ref{sec:more-metrics}).
However, while both algorithms converge to almost the same optical quality, the DL algorithm requires an extra iteration to converge.
The implications of this extra step are discussed in Section~\ref{sec:discussion}.

\subsection{Estimating the wavefront using a single donut}
\label{sec:single-donut}

\begin{figure*}
    \centering
    \includegraphics{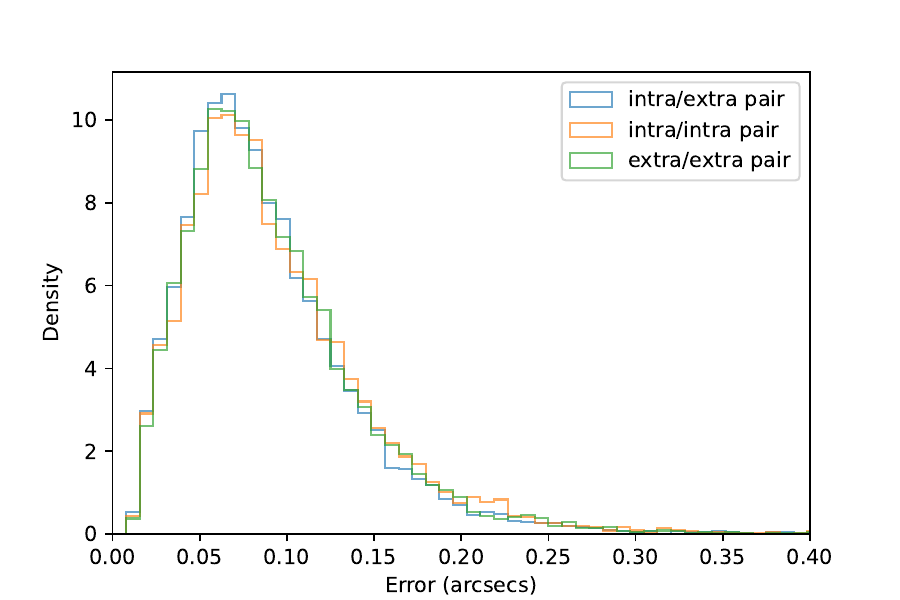}
    \caption{
        Distribution of wavefront estimation errors for the DL model, using intra/extra, intra/intra, and extra/extra pairs.
        These distributions coincide, indicating that the DL model can accurately estimate the wavefront only using information from one side of focus.
    }
    \label{fig:ml-pairs}
\end{figure*}

In the preceding sections, we used the DL model to estimate the wavefront using a pair of intra- and extra-focal donuts, enabling an apples-to-apples comparison with the TIE solver.
The DL model, however, can estimate the wavefront using only a single donut, meaning it does not need information from both sides of focus.
To evaluate DL performance on one side of focus, we calculate wavefront estimation errors while averaging the wavefront from pairs of donuts \emph{on the same side of focus}.
We do this pairing to make sure we are still getting the benefit of noise reduction from averaging a pair of wavefronts, while only using information from one side of focus.
You can see in Fig.~\ref{fig:ml-pairs} the DL model does well with \emph{only} intra-focal or extra-focal images.
Interestingly, it appears that averaging two wavefronts only reduces the noise by a factor of $\sim 1.2$, $15\%$ less than the factor of $\sqrt{2} \sim 1.41$ expected for uncorrelated Gaussian error.
This is due to correlation of atmospheric errors across individual CWFSs due to low-altitude turbulence.

The ability to estimate the wavefront using only images from one side of focus has several advantages.
For example, in crowded fields, you may not be able to find an intra-/extra-focal pair where each donut is sufficiently bright, and is not too heavily blended or vignetted. 
It removes the need to compensate for differences between the two sources in a pair, which the baseline algorithm must do in order to apply the TIE.
It can also provide more localized wavefront estimates across each CWFS, which provides more leverage when interpolating the wavefront across the full focal plane.
We note that the methods of \citet{tokovinin2006}, \citet{janish2012}, and \citet{roodman2014} are all also capable of estimating the wavefront using an image from only one side of focus.

There may also be downsides, however, to using information from a single side of focus.
Distortion of the pupil, attenuation of the mirror, and differential chromatic refraction (DCR) all have effects that are independent of the side of focus on which a donut is imaged.
Wavefront aberrations, on the other hand, change parity as they pass through focus.
For example, astigmatism and DCR both elongate sources, but the former rotates by 90\deg when you pass through focus while the latter is always oriented towards zenith.
Using information from both sides of focus allows you to distinguish these effects.
We tested the impact of DCR by simulating very blue stars at airmass 4, imaged in the Rubin $g$ band, while toggling the simulated DCR on and off.
We find that in this extreme case, DCR increases the median wavefront estimation error by 7\% when using information from only one side of focus, while the median only increases by 5\% when averaging estimates from both sides of focus.
This is only a modest increase, even in a situation where DCR will be the most severe.
Furthermore, if this does become a limiting factor in certain circumstances, the effects of DCR can be predicted and separately modeled.

\section{Discussion \& Conclusion}
\label{sec:discussion}

In this paper, we have built a deep learning (DL) model for wavefront estimation.
Previous works have built prototypes for this application \citep{thomas2020a, thomas2021, yin2021}, but this is the first time such a system has been integrated into the Rubin AOS pipeline, and validated against the baseline algorithm.
We have demonstrated that under ideal conditions, where the images do not have bright blends and are not vignetted by the camera, the DL system outperforms the baseline model, improving the median wavefront estimation error by a factor of 2.
In fact, the estimation error for the DL model is comparable to the expected minimum error due to random phase fluctuations from the atmosphere.
This indicates that the DL system is performing as good as can be expected when given high-quality data.

Furthermore, the DL system is far more robust than the baseline algorithm when presented with less ideal data.
When the intra-focal donut is too far from the center of the focal plane, its edge is sharply vignetted by the camera body.
This results in an exponential decrease in accuracy for the baseline algorithm, while the DL model is unaffected.
Furthermore, blending of the central source with neighboring sources greatly diminishes the accuracy of the baseline method, and often causes catastrophic failures.
Again, the DL model is largely unimpacted, improving the median wavefront estimation error by a factor of 14.
Both algorithms begin to degrade when the total SNR of the source drops below 200, but the degradation of the DL algorithm is less severe.

In addition to the strengths listed above, the DL model can accurately estimate the wavefront using only a single image, whereas the baseline algorithm requires a pair of intra- and extra-focal images.
The DL model is also approximately 40 times faster when, like the baseline algorithm, it is applied to pairs one-at-a-time, but much greater speed-ups are achievable if the DL model is applied to many images simultaneously.

These strengths make the DL model a valuable addition to the Rubin AOS.
It will allow the AOS closed loop to operate in the densest 8\% of fields where the baseline algorithm fails due to the inability to select a pair of unvignetted and unblended sources.
This will expand by 1400~deg$^2$ the area useful for precision galactic science that targets the crowded plane of the Milky Way.
Higher performance with degraded data and the ability to estimate the wavefront given a single source from each CWFS also increases the robustness of the AOS to a wide variety of conditions, such as satellite trails and failing CCDs.

The one test where the baseline algorithm outperforms the DL model is the convergence rate of the AOS closed loop simulated with PhoSim.
PhoSim is a different raytracer than that used to simulate the training data for the DL model.
The DL model still meets the closed loop requirements and converges to the same optical quality, but requires an additional closed loop iteration to do so.
This might be due to chance, and simulating many more closed loops in a wide variety of conditions may show that the DL model exceeds the baseline on this test as well.
\textbf{
However, these results might also indicate a need to simulate a wider range of wavefront perturbations, as the initial perturbation in the closed loop might lie outside the area of parameter space covered by our training set.
}

To prepare for commissioning, we plan to generate a larger library of simulations.
These simulations will more densely and expansively sample the space of potential optical perturbations, as well as include a larger number of independently simulated atmospheres.
This will reduce the dangers of overfitting and network extrapolation.
We also plan to include more realistic estimates for stellar temperatures in order to test for chromatic effects, and to upgrade our simulations to include more image effects, such as mirror print through, diffraction from the telescope support structure, cosmic rays, and bad pixels.

We expect that domain adaptation, the technique of adapting a model trained on one data set so that it performs well on a different target data set, will also be necessary for commissioning our deep learning system.
This is because real data is always messy in ways not captured by simulations, and deep learning systems need to develop robustness to these differences in order to perform accurate inference on real data.
This task will be difficult for the Rubin AOS, since the vast majority of Rubin images will be ``unlabeled'' in the sense that the true wavefront will not be known\footnote{The exception to this will be a small number of exposures for which we will apply large, known perturbations to the telescope control parameters, so that the applied perturbations will dwarf any natural, unknown perturbations.}.
For this reason, we expect strategies such as domain adversarial training \citep{ciprijanovic2021, ciprijanovic2022} will be valuable when commissioning this system on real data.
Domain adversarial training uses a discriminator network that takes the CNN image features as input, and tries to guess whether the original image was real or simulated.
By training the CNN to outsmart the discriminator, the CNN learns to ignore artifacts and features that are unique to the real images, and instead focus on the physical information present in both the real images and the simulated training set.

It may also be advantageous to further optimize the network architecture for domain adaptation, since two of our architectures achieved similar performance on the Batoid test set, but differing performance in the PhoSim closed loop.
Overcoming the obstacle of domain adaptation will be the greatest challenge during commissioning.
We have begun working towards these goals using data from the Rubin Observatory Auxiliary Telescope (AuxTel), which is a simpler system and is already producing data.

Finally, the results presented in Section~\ref{sec:metrics} will also be useful in improving the baseline wavefront estimation algorithm.
Using the wavefront estimation error as a function of SNR, vignetting, blend distance, and the number of faint blends will help refine the donut selection criteria.
The evidence of a model bias compared to the atmospheric uncertainty and the DL performance motivates additional investigation into the modeling assumptions of the baseline algorithm.

\section*{Acknowledgments}

We would like to thank David Thomas who provided some scaffolding for early stages of this project, Tam\'as Budav\'ari and Yashil Sukurdeep for helpful feedback throughout, Aaron Roodman for feedback on a proposal related to this work, Aleksandra \'Ciprijanovi\'c for discussions related to domain adaptation and commissioning, Will Sutherland for his insights on atmospheric noise correlations, and Chris Stubbs for reviewing the manuscript.

J.F.C is supported by the U.S. Department of Energy, Office of Science, under Award DE-SC0011665.
J.E.M. was supported by the U.S. Department of Energy under contract number DE-AC02-76SF00515.
This material is based upon work supported in part by the National Science Foundation through Cooperative Agreement AST-1258333 and Cooperative Support Agreement AST-1202910 managed by the Association of Universities for Research in Astronomy (AURA), and the Department of Energy under Contract No. DE-AC02-76SF00515 with the SLAC National Accelerator Laboratory managed by Stanford University.
Additional Rubin Observatory funding comes from private donations, grants to universities, and in-kind support from LSSTC Institutional Members.

This work has made use of data from the European Space Agency (ESA) mission \emph{Gaia} processed by the \emph{Gaia} Data Processing and Analysis Consortium (DPAC).
Funding for the DPAC has been provided by national institutions, in particular the institutions participating in the \emph{Gaia} Multilateral Agreement.

\section*{Software}

The simulation \citep{donutsims_repo} and neural network \citep{mlaos_repo} code is stored on GitHub and Zenodo.
The data and trained model are also stored on Zenodo \citep{mlaos_paper_deposit}.
Wavefront estimation was performed with the Rubin Wavefront Estimation Pipeline, on GitHub at \url{https://github.com/lsst-ts/ts_wep}.
The Rubin Active Optics closed loop was simulated using code stored on GitHub at \url{https://github.com/lsst-ts/ts_phosim}.
This repository depends on the software for the Rubin Optical Feedback Controller, on GitHub at \url{https://github.com/lsst-ts/ts_ofc}.
The plots in this paper were created using the code on GitHub at \url{https://github.com/jfcrenshaw/aos_notebooks}.

In addition to the repositories described above, this paper made use of the following software:
Batoid \citep{batoid},
GalSim \citep{galsim},
Jupyter \citep{jupyter},
Matplotlib \citep{matplotlib},
NumPy \citep{numpy},
Pandas \citep{pandas, pandas-software},
PhoSim \citep{phosim},
PyTorch \citep{pytorch},
PyTorch Lightning \citep{pytorch-lightning},
ResNet-18 \citep{resnet},
Rubin OpSim \citep{opsim},
SciPy \citep{scipy}.

\appendix

\section{Clarifying the LSST Optical Error Budget}
\label{sec:error-budget}

Throughout this paper, there are several references to the LSST error budget.
The numbers cited are closely related, but subtly different.
In this section we clarify the relationship between these numbers.
Note that when combining errors from multiple sources, PSF FWHM values are meant to be added in quadrature and PSSN values are meant to be multiplied.

In Section~\ref{sec:intro}, we state:
\begin{displayquote}
    The median seeing at Rubin's site on Cerro Panch\'on is 0.65" \citep{ivezic2019}, and Rubin's optical system is required to degrade this value by no more than 0.4".
\end{displayquote}
This 0.4" requirement, from \citet{LSE-29}, is for the entire optical system, including all contributions from both the telescope and camera.
In particular, of this 0.4" requirement, 0.25" is allocated to the telescope, 0.3" is allocated to the camera, and 0.08" is allocated to the optical design.

Later in this same section, we state:
\begin{displayquote}
    To deliver the optical quality required for LSST, the AOS must limit the PSF (Point Spread Function) FWHM (Full Width Half Maximum) contribution of optical aberrations to no more than 0.09", including only 0.079" due to misestimation of the wavefront -- less than 10\% of the total error budget. 
\end{displayquote}
The 0.09" requirement is the portion of the telescope error budget allocated to the active optics system, and of this, 0.079" is allocated to optical errors due to misestimation of the wavefront.
These numbers are sourced from \citet{LTS-186} and \citet{LTS-124}.

In Section~\ref{sec:closedLoop}, we state:
\begin{displayquote}
    The AOS closed loop is required to converge to a FWHM of less than 0.19" and a PSSN greater than 0.91.
\end{displayquote}
The 0.19" requirement cited here contains the 0.09" allocated to the active optics, plus additional errors of 0.167" allocated to M1M3 and 0.032" allocated to M2 (which are also captured by the closed loop simulation).
The quoted PSSN (0.91) is similarly the product of the PSSN specified for the active optics (0.981), M1M3 (0.933), and M2 (0.998).
These numbers are also sourced from \citet{LTS-124}.

\section{Sampling Rubin optical perturbations}
\label{sec:perturbations}

\begin{figure*}[b]
    \centering
    \includegraphics[scale=0.85]{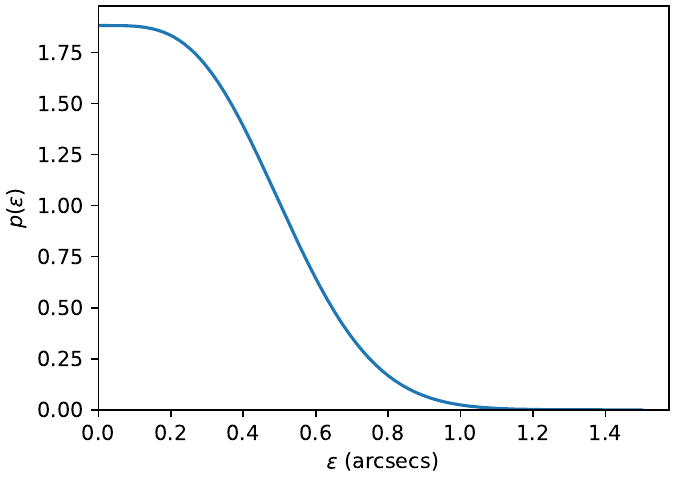}
    \caption{The distribution of PSF FWHM degradations sampled for the simulations.}
    \label{fig:psf-distribution}
\end{figure*}

For each simulated telescope pointing, we randomly perturb the 50 parameters that specify the optical alignment and the figure of the mirror.
Parameters 1-5 and 6-10 describe the rigid body motions of M2 and the camera, respectively.
These include translations along all three axes (in microns), and rotations about the $x$ and $y$ axes (in arcseconds).
Parameters 11-30 and 31-50 are the bending mode amplitudes in microns for M1M3 and M2, respectively.

To perturb these parameters, we first sample a PSF FWHM degradation from a folded Gaussian distribution whose standard deviation is equal to its mean (i.e., $\sigma = \mu$).
This distribution was heuristically chosen to yield a distribution that emphasizes small degradations, but has a tail towards much larger degradations.
The PDF of this distribution is
\begin{align}
    p(\varepsilon) = \frac{1}{\sqrt{2\pi \mu^2}} \left(
        e^{-(\varepsilon/\mu - 1)^2/2} + e^{-(\varepsilon/\mu + 1)^2/2}
    \right),
    \label{eq:psf-distribution}
\end{align}
where $\varepsilon$ is the PSF FWHM degradation in arcseconds.
This distribution is plotted in Fig.~\ref{fig:psf-distribution}.
The parameter $\mu$ is chosen such that $\langle \varepsilon \rangle = 0.3"$, using the formula
\begin{align}
    \mu = \langle \varepsilon \rangle \left[
        \sqrt{\frac{2}{\pi}} e^{-1/2} + \mathrm{erf}\left( \frac{1}{\sqrt{2}} \right)
    \right]^{-1}.
\end{align}
In other words, we sample $\varepsilon$ from a Gaussian with mean and standard deviation equal to $\mu$, then take the absolute value.
This determines the total PSF FWHM degradation associated with the perturbation.
We then determine what portion of this PSF degradation to assign to each telescope parameter by sampling from the 50-dimensional hypersphere with radius $\varepsilon$.
Finally, we use the values of $\Delta \mathrm{PSF} / \Delta \mathrm{perturbation}$ determined in Appendix~\ref{sec:psf-conversion} to convert these PSF degradations to parameter perturbations in the appropriate unit.

We use Batoid to simulate images from telescopes with these simulated perturbations. 
For each simulation, we use Batoid to extract the wavefront and calculate the corresponding Zernike coefficients.
As described in Section~\ref{sec:wfest}, these wavefronts are at the center of each CWFS, for a fiducial wavelength of 1~$\mu$m, and have the intrinsic aberrations of the telescope design subtracted.

\section{Estimating PSF FWHM degradation} 
\label{sec:psf-conversion}

\begin{figure*}[b]
    \centering
    \includegraphics[scale=0.81]{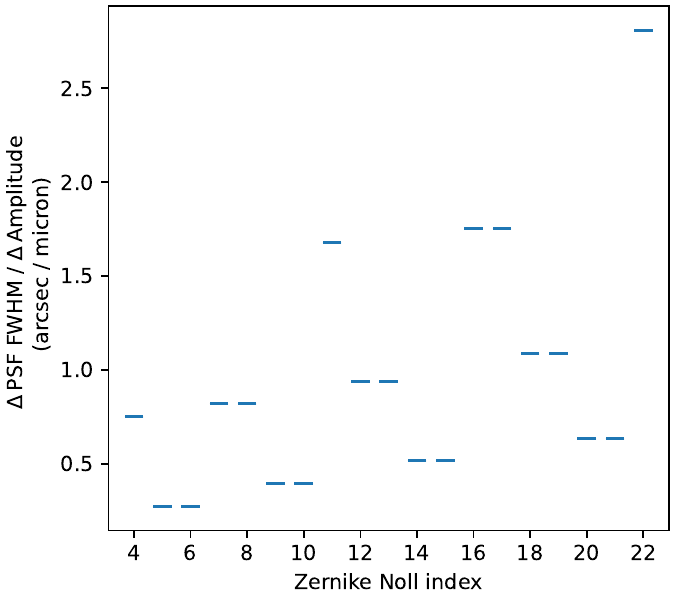}
    \caption{
        The PSF FWHM weighting factors, $s_i$ for the Rubin Observatory.
    }
    \label{fig:fwhm-scale-factors}
\end{figure*}

\begin{figure*}
    \centering
    \includegraphics{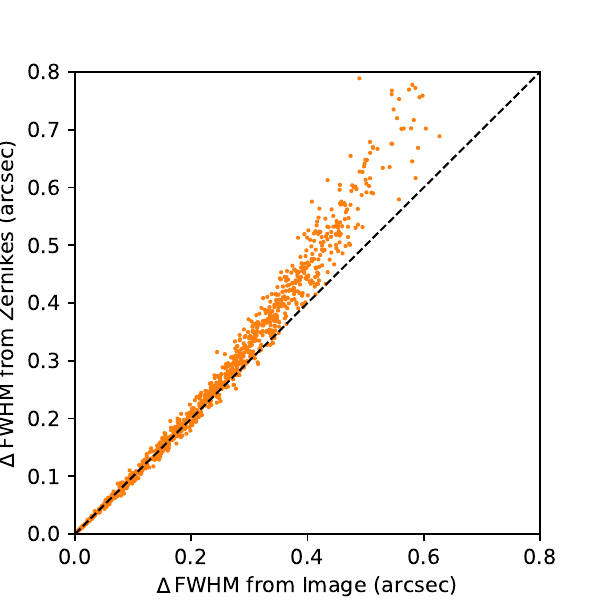}
    \caption{
        Comparing the increase in PSF FWHM due to wavefront aberrations for 1000 simulated stars with random Zernike amplitudes.
        The $x$-axis is the increase in FWHM as measured from the simulated images, while the $y$-axis is the value estimated from our weighted Zernike scheme.
        The dashed line indicates perfect correspondence.
        The two methods agree for FWHM increases below 0.25".
        Above this level, the weighted Zernike scheme overestimates the increase in FWHM.
    }
    \label{fig:fwhm-sim}
\end{figure*}

It is useful to express wavefront estimation errors in a unit that is readily interpretable in terms of the optical quality.
In this appendix, we derive a set of scaling factors, $s_i$, such that a given wavefront perturbation $\varphi = \sum_i \alpha_i Z_i$ results in an increase of the PSF FWHM by the amount
\begin{align}
    \text{PSF FWHM increase} = \sqrt{ \sum_i (s_i \alpha_i)^2 }.
\end{align}

To derive these factors, recall that the gradient of the wavefront provides a map from pupil position to focal plane position (e.g., \citealt{roddier1993}).
We can express the gradients of the wavefront with two more Zernike expansions:
\begin{align}
    \frac{\partial \varphi}{\partial x} = \sum_i \alpha^{x}_i Z_i
    \quad \quad
    \frac{\partial \varphi}{\partial y} = \sum_i \alpha^{y}_i Z_i,
\end{align}
the coefficients of which can be determined analytically from the expansion coefficients of $\varphi$ \citep{noll1976,zhao2007,stephenson2014} or calculated using the Galsim Zernike library \citep{galsim}.
The variance of the $x$ position of photons on the focal plane is then
\begin{align}
    \sigma_{x}^2
    = \frac{1}{A} \int d^2\!\rho \left[ \frac{d\varphi}{dx}(\rho) \right]^2
    &= \frac{1}{A} \sum_{i,j} \alpha_i^x \alpha_j^x \int d^2\!\rho \,  Z_i(\rho) Z_j(\rho) \\
    &= \sum_{i,j} \alpha_i^x \alpha_j^x \, \delta_{ij}
    = \sum_i (\alpha_i^x)^2,
\end{align}
where $\rho$ is a 2D position vector on the pupil and we use the Zernike normalization defined in Eq.~\ref{eq:coefficients}.
The variance in $y$ positions is calculated analogously.
Fitting an isotropic Gaussian to the photon positions on the focal plane yields the variance
\begin{align}
    \sigma^2 
    = \frac{1}{2} (\sigma_x^2 + \sigma_y^2) 
    = \frac{1}{2} \sum_i (\alpha_i^x)^2 + (\alpha_i^y)^2.
\end{align}
Finally, the FWHM can be calculated with the regular scaling for a Gaussian:
\begin{align}
    \mathrm{FWHM} = 2 \sqrt{2 \ln 2} \, \sigma
\end{align}
Scale factor $s_i$ can be calculated in units of arcseconds per micron by applying the above formulae to the wavefront $\varphi = \alpha_i Z_i$, where $\alpha_i = 1~\mu\text{m}$.
The values of $s_i$ for the Rubin Observatory are shown in Fig.~\ref{fig:fwhm-scale-factors}.
All of the scale factors are of $\mathcal{O}(1)$.
Note that the values of these scale factors depend on the geometry of the pupil, and are therefore telescope specific.

We used simulations to validate our Zernike-to-PSF-FWHM conversion scheme.
First we simulated a pair of in-focus stars: one with an unaberrated wavefront, and one with random aberrations in Zernikes 4 to 23.
The random aberrations were independently drawn from a Gaussian distribution --- i.e. the aberrations in each mode have identical variance, but no covariance between different Zernike modes.
For each pair of stars, we used Galsim to measure the PSF FWHM, and took the difference to estimate the increase to the FWHM due to the wavefront aberrations.
We then calculated the expected FWHM increase from the Zernike amplitudes, using the method described above.

The results of this experiment are shown in Fig.~\ref{fig:fwhm-sim}.
For FWHM degradations $\lesssim 0.25$~arcsecs, the Zernike conversion scheme described in this appendix is a good match to the degradations measured from image simulations.
Above this level, our scheme mildly overestimates the FWHM degradation, likely due to degeneracies (and therefore cancellations) in the way each Zernike mode impacts the PSF FWHM, which are not captured by our method.
This is not a significant problem as Fig.~\ref{fig:ideal} shows the majority of our wavefront errors correspond to PSF FWHM degradations below the 0.25 arcsecond level.
It can be assumed that the errors in the high error tail of Fig.~\ref{fig:ideal} are slightly over estimated.
This caveat, however, does not change the conclusions of the main text.

Note that while the majority of Zernike estimation errors lie below the 0.25 arcsecond level, the same is not necessarily true of the Zernike amplitudes themselves.
It is still useful to plot Zernike amplitudes in units of PSF FWHM contribution, however, as this enables a by-eye estimate of how the Zernike estimation residuals impacts the PSF of the telescope.

Similarly to the Zernike amplitude conversions described above, you can calculate PSF FWHM degradation for each of the optical parameters discussed in Appendix~\ref{sec:perturbations}.
We perturb each parameter 1 unit, use Batoid raytracing to retrieve the corresponding wavefront, and then proceed as above.

\section{Simulating the atmosphere} 
\label{sec:atm}

\begin{figure*}[b]
    \centering
    \includegraphics[scale=0.85]{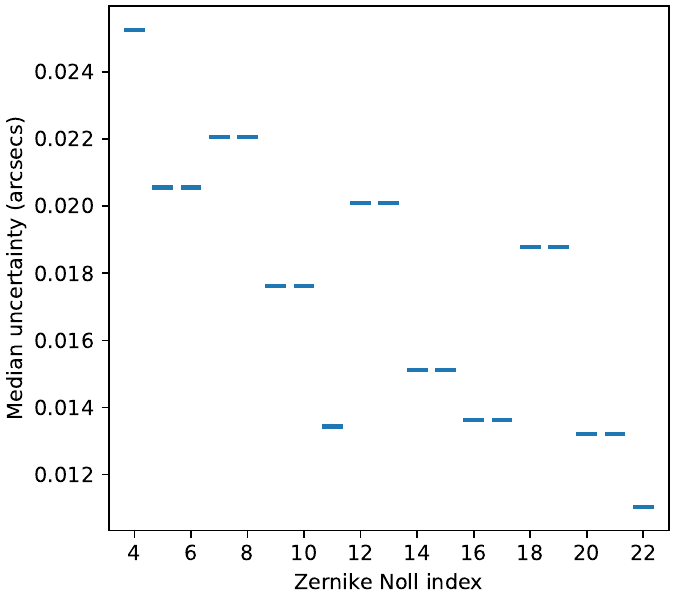}
    \caption{
        Median $1\sigma$ error due to the atmosphere for each Zernike coefficient $\alpha_i$.
    }
    \label{fig:median-atm}
\end{figure*}

In addition to distortions from misalignment and mirror deformation, the Rubin optical wavefront is distorted by atmospheric turbulence.
Because Rubin's field of view is so large, we cannot correct for this turbulence.
In effect, the atmosphere adds a random Zernike series to the Zernike series we are trying to estimate.
These random phase variations from the atmosphere are an irreducible noise floor for wavefront estimation, so it is useful to understand their statistics.

Using Eq.~\ref{eq:coefficients}, we can write the covariance of the coefficients $\alpha_i$ as
\begin{align}
    \langle \alpha_i \alpha_j \rangle 
    = \frac{1}{A^2} \int d^2\!\rho \int d^2\!\rho' ~ Z_i(\rho) Z_j(\rho') \, \langle \varphi(\rho) \varphi (\rho') \rangle.
    \label{eq:cov}
\end{align}
We assume the von K\'arm\'an turbulence model \citep{vonkarman1948}, in which turbulence begins at an upper scale $L_0$ and cascades down to the viscous dissipation scale $\ell_0$.
The phase covariance for von K\'arm\'an turbulence is \citep{carlsten2018}
\begin{align}
    \langle \varphi(\rho) \varphi (\rho') \rangle = 
    0.0858 \left(\frac{L_0}{r_0}\right)^{5/3} \left(\frac{2\pi r}{L_0}\right)^{5/6} K_{5/6}\left(\frac{2\pi r}{L_0}\right),
    \label{eq:von-karman}
\end{align}
where $r \equiv |\rho - \rho'|$, and $K$ is a modified Bessel function of the second kind.
$r_0$ is the usual Fried parameter \citep{fried1965}, which, for von K\'arm\'an turbulence, is related to $L_0$ and the PSF FWHM via the formula \citep{tokovinin2002}
\begin{align}
    \mathrm{FWHM} = \frac{0.976 \lambda}{r_0} \sqrt{1 - 2.183 \left(\frac{r_0}{L_0}\right)^{0.356}}.
\end{align}

We assume individual turbulent cells do not evolve during the 15 second exposures, and so the coherence time of turbulence is determined by the wind speed, which moves new turbulent cells across the field of view during an exposure.
The coherence time is approximately $\tau_0 = 0.31 \, r_0 / v_e$, where $v_e$ is the average wind speed, weighted by the strength of turbulence in each layer \citep{roddier1981}.
Thus, during a 15 second exposure, the phase covariance is reduced by a factor of $N = 15\,s / \tau_0$.

We use GalSim \citep{galsim} to simulate an atmosphere with von K\'arm\'an turbulence in 6 layers.
We use the turbulence profile measured on Cerro Armazones \citep{ellerbroek2008} (with 10\% Gaussian noise).
Outer scales $L_0$ are randomly drawn from a log-normal distribution with mean 30 m and standard deviation 20 m, truncated between 10 and 100 m.
Wind speeds are isotropically sampled with a uniform distribution between 0 and 20 m/s.
We assume constant air pressure of 69~kPa, H$_2$0 pressure of 1~kPa, and temperature of 293~K.

To estimate the impact of turbulence on Zernike estimation, we numerically evaluate Eq.~\ref{eq:cov} while sampling the PSF FWHM and wavelength from the baseline LSST OpSim simulation, and $L_0$ and $v_e$ from the distributions described above.
The median standard deviation of each Zernike coefficient is shown in Fig.~\ref{fig:median-atm}.
The sum in quadrature of these values is 0.072".
For a comparison of the wavefront estimation algorithm errors to the atmospheric error, see Section~\ref{sec:metrics-ideal}.

\citet{noll1976} studied the projection of atmospheric statistics onto Zernike coefficients for a circular aperture (i.e., no central obscuration) and Kolmogorov turbulence (the $L_0 \to \infty$ limit of Eq.~\ref{eq:von-karman}; \citealt{fried1965}).
Their study found that for each coefficient, the variance scales like $r_0^{-5/3}$, and that the total variance contained by all coefficients greater than Noll index $j$ scales like $j^{-\sqrt{3}/2}$ for large values of $j$.
Despite studying von K\'arm\'an turbulence for an aperture with a large central obscuration, we find the same scaling relations (for fixed $L_0$, before dividing by $N$).
We note that the variance averaged over a 15 second exposure scales like $r_0^{-2/3}$, as this entails a division by $N \propto r_0^{-1}$.

\section{Additional wavefront estimation metrics}
\label{sec:more-metrics}

\begin{figure*}[b]
    \centering
    \includegraphics[scale=0.85]{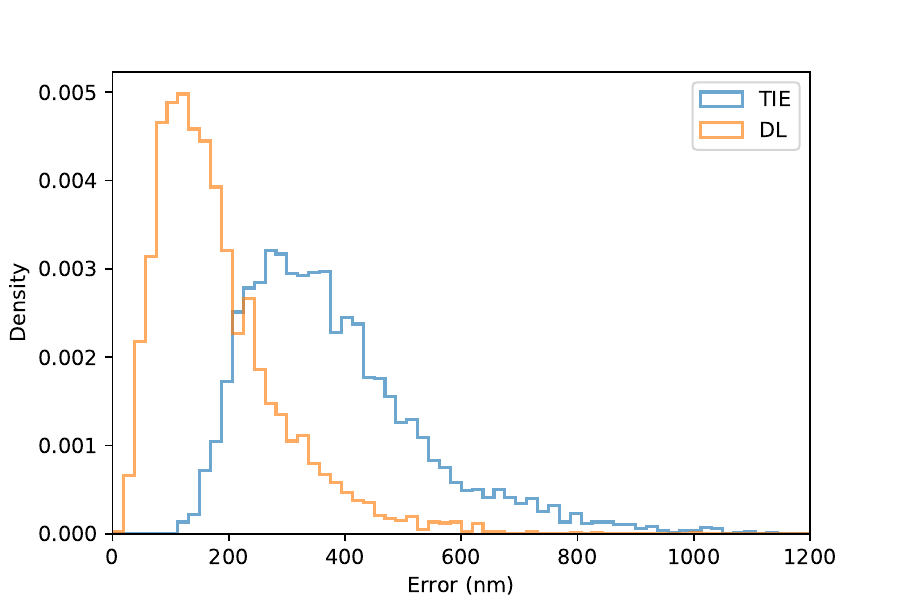}
    \caption{
        Distribution of wavefront estimation errors for the DL and TIE methods in ideal conditions (no camera vignetting and no bright blends), in units of nanometers.
        This is a near duplicate of Fig.~\ref{fig:ideal}, except with a different unit on the $x$-axis.
    }
    \label{fig:ideal-nm}
\end{figure*}

\begin{figure*}
    \centering
    \includegraphics[scale=0.8]{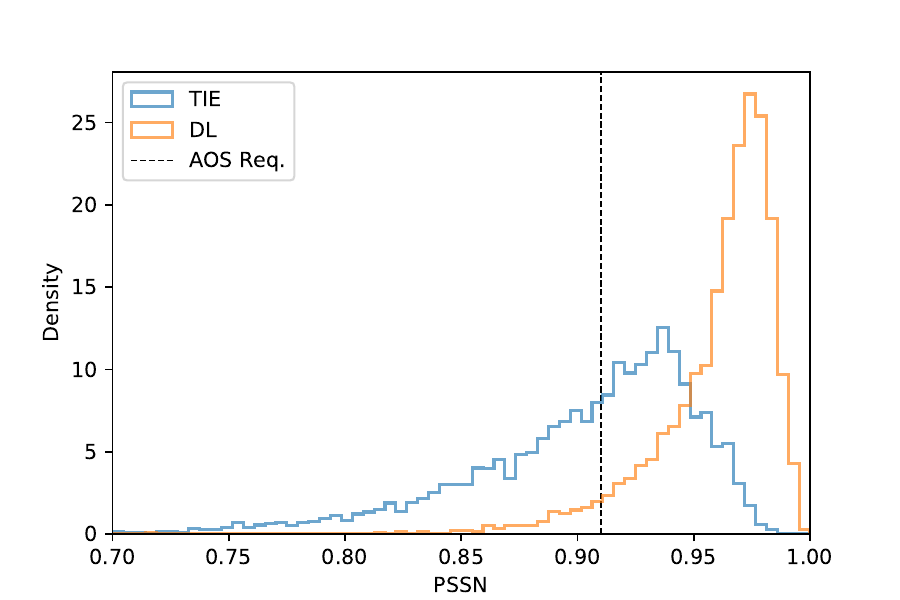}
    \caption{
        Distribution of PSSN values after correcting for aberrations using the estimated wavefront from the DL and TIE methods.
        The vertical black line is the AOS wavefront estimation error requirement of 0.91.
    }
    \label{fig:ideal-pssn}
\end{figure*}

\begin{figure*}
    \centering
    \includegraphics[width=0.85\textwidth]{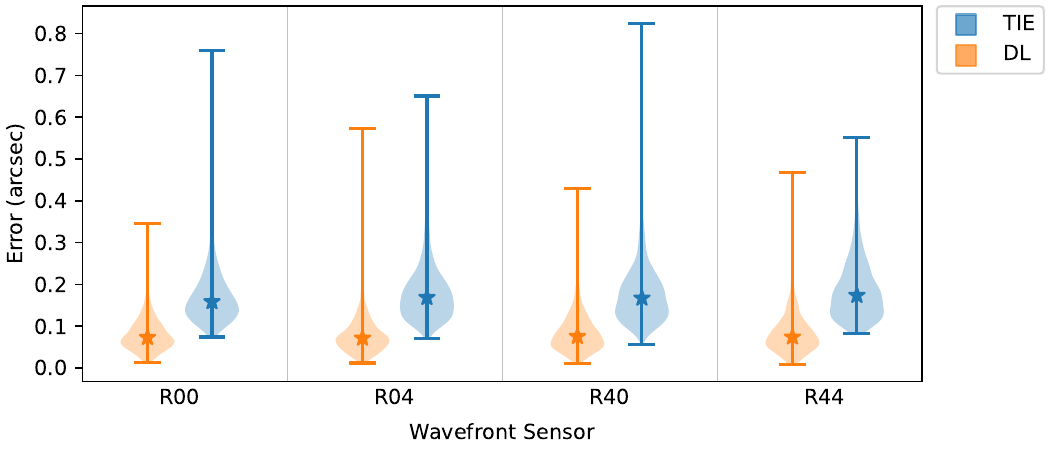}
    \caption{
        Distribution of wavefront estimation errors for each CWFS, labeled according to the Rubin convention.
        The stars denote the median error.
    }
    \label{fig:sensor-violin}
\end{figure*}

\begin{figure*}
    \centering
    \includegraphics[width=0.85\textwidth]{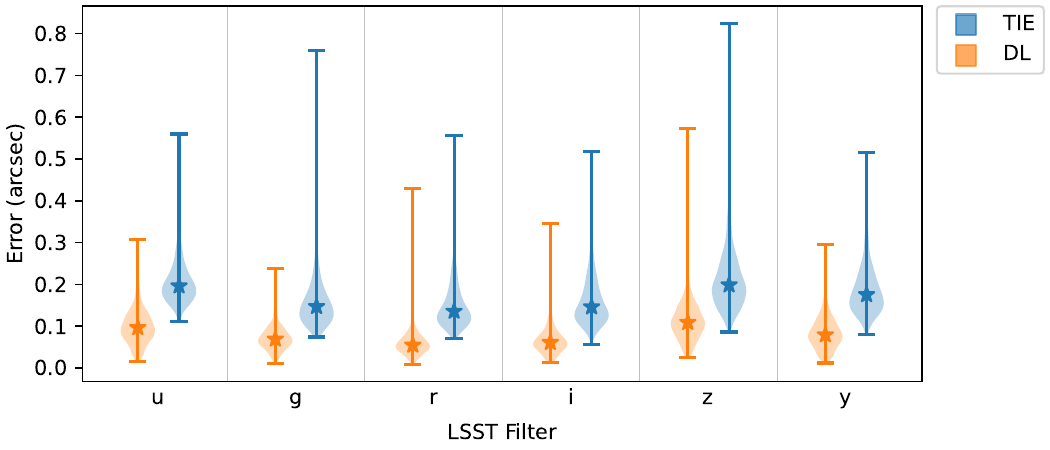}
    \caption{
        Distribution of wavefront estimation errors for each of the LSST filters.
        The stars denote the median error.
    }
    \label{fig:filter-violin}
\end{figure*}

\begin{figure*}
    \centering
    \includegraphics[width=0.85\textwidth]{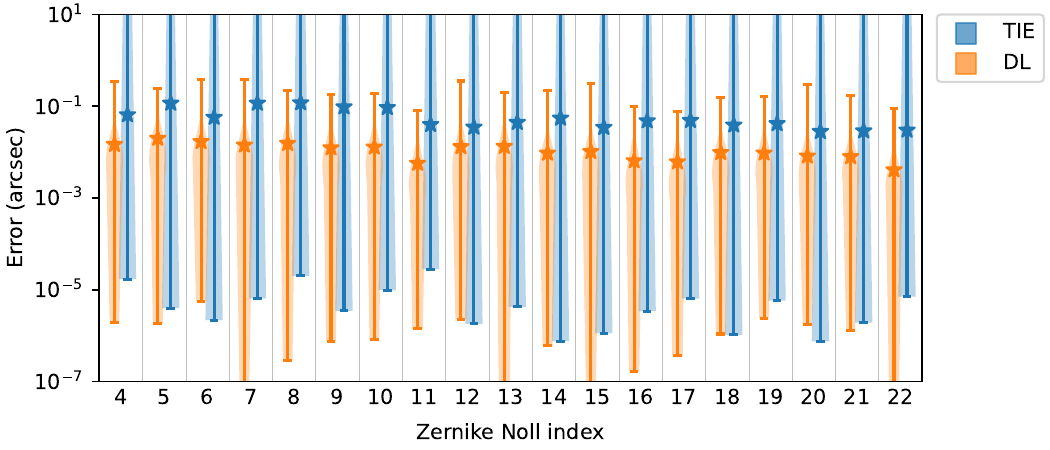}
    \caption{
        Distribution of errors for each Zernike coefficient, $\alpha_i$.
        The stars denote the median error.
    }
    \label{fig:zernike-violin}
\end{figure*}

In this appendix, we provide a few more figures to check that the conclusions of this paper do not depend on the choice of PSF FWHM as the metric for wavefront estimation, as well as a few figures that check the consistency of our results when selecting different subsets of the test set.

Fig.~\ref{fig:ideal-nm} is a near duplicate of Fig.~\ref{fig:ideal}, except that the wavefront residuals are measured in nanometers.
This provides a check on our conversion to PSF FWHM by showing that even without this weighting scheme, the DL method outperforms the TIE.
The shapes of the histograms in these two plots are nearly identical, indicating that the PSF FWHM carries nearly the same information as the residual in nanometers, with the benefit of having a unit that is more readily interpretable in terms of optical quality.

Fig.~\ref{fig:ideal-pssn} is another histogram comparing the performance of the DL and TIE methods on pairs of stars (with no camera vignetting or bright blends).
This figure shows PSSN values after correcting for simulated aberrations using the estimated wavefront from each method.
Recall that PSSN ranges between zero and one, with greater values indicating better optical quality.
This figure shows that the DL method also outperforms the TIE method on maximizing the PSSN, providing a check that training to optimize the PSF FWHM does not come at the expense of other optical quality metrics.
Interestingly, both the DL and TIE methods achieve the PSSN required by the error budget (marked with the dashed line) using a single pair of donuts more often than they achieve the required PSF FWHM (cf. Fig.~\ref{fig:ideal}).

Fig.~\ref{fig:sensor-violin} provides a check that both methods have similar performance on each of the wavefront sensors.
There is no reason to expect different performance on different sensors, so this plot provides a consistency check on the simulations and the training of the DL method.

Fig.~\ref{fig:filter-violin} shows the performance for each LSST filter.
The errors are comparable between every band, with slightly worse performance for the $u$, $z$, and $y$ bands.
This is expected as the average SNR in these bands is slightly worse.

Fig.~\ref{fig:zernike-violin} compares the error distribution for each Zernike coefficient.
This shows that the DL method outperforms the TIE method on wavefront estimation for every Zernike mode.
In addition, compared to the TIE, the DL method has less variation in performance for different Zernike modes.


\newpage

\bibliography{references}   
\bibliographystyle{aasjournal}   

\end{document}